\documentclass[prb,aps,epsf,floats,twocolumn]{revtex4}
\usepackage[cp1251]{inputenc}
\usepackage[english]{babel}

\input epsf

\begin{document}
\title{ Influence of free carriers on exciton ground states in quantum wells }

\author{A.A. Klochikhin $^{1,2}$}
\email{albert.klochikhin@mail.ioffe.ru}
\author{V.P. Kochereshko$^{1}$ }
\author{ S. Tatarenko$^3$}
\address{$^{1}$Ioffe
Physico-Technical Institute, 194021 St. Petersburg, Russia\\
$^{2}$ Nuclear Physics Institute, 350000 St. Petersburg, Russia\\
$^3$CEA-CNRS group ''Nanophysique et Semiconducteurs,'' Institut
N$\acute{e}$el, CNRS and Universite  Joseph Fourier, 25 Avenue des
Martyrs, 38042 Grenoble, France }

\date{Received \today}

\begin{abstract}
{ The influence of free carriers on the ground state of the
exciton at zero magnetic field in a quasi-two-dimensional quantum
well doped with electrons is considered in the framework of the
random phase approximation. The effects of the
exciton-charge-density interaction and the inelastic scattering
processes due to the Hartree-Fock electron-electron exchange
interaction are taken into account.  The effect of phase-space
filling is considered using an approximate approach. The results
of the calculation are compared with the experimental data
available.}
\end{abstract}

\maketitle

PACS number(s): 71.10.Ca, 71.45.Gm, 73.21.Fg, 78.55.Et

\section{Introduction}
The presence of free electrons in quasi-two-dimensional  semiconductor structures
strongly alters their physical characteristics. In optical
investigations of these objects, the photon excites a conduction
electron -- valence hole pair. To understand the optical properties, it is crucially important to take into account
the interactions between this pair and the free carriers.

At low and intermediate electron concentrations, the Coulomb
interaction between the electron and the hole leads to formation
of the bound exciton  state. This shrinks the observable band gap
$E_G$ by the exciton binding energy $E_B$. Similarly to degenerate
3D semiconductors, the electron-electron interaction causes a
static screening of the electron-hole Coulomb potential. The
screening as well as the filling of phase-space up to the Fermi
energy leads to a strong decrease of the exciton binding energy as
free electron concentration  increases. At some concentration of
free electrons the exciton bound state disappears and the
so-called Mahan exciton can be expected. Considerable attention
has been paid in the literature to experimental and theoretical
investigations of the 2D-systems  at high concentration of free
carriers.
 The theory of this phenomenon was developed
in Ref.\cite{Mahan} for the 3D crystal and in
Ref.\cite{Schmitt-Rink-1,Schmitt-Rink-2,Hawrylak,Bauer} for 2D
systems.

Experimental data\cite{Cox1,Cox2,Cox3,Teran,Kukushkin}  give
evidence that, within a relatively narrow interval  of free
electron concentrations below the metallic phase, the optical
absorption band associated with the bound exciton shifts and
broadens, and its oscillator strength decreases as carrier
concentration is increased. The dependence of these
characteristics on the electron concentration allows one to relate
these processes with exciton-electron interactions.

 In addition, starting from very low concentrations
 of free electrons in the quantum well,
 a bound trion (negatively charged exciton)
 state\cite{Combescot1,Combescot2,Combescot3,Combescot4} is observed
 in experiments\cite{Cox1,Cox2,Cox3,Teran,Kukushkin}.
 The present article will not be concerned
 with properties related to the trion state, but
 its existence provides another channel for exciton-electron
 interactions.

The observed broadening of the exciton
line\cite{Cox1,Cox2,Cox3,Teran} clearly shows that not only static
screening but also inelastic dynamical processes should be
considered for the explanation of the exciton spectrum.

The exciton created by a photon has almost zero wave-vector.
Further interaction of the exciton with free electrons followed by
scattering  of  a conduction electron near the Fermi level with
$p\approx p_F$ creates a conduction electron -- conduction hole
pair. This pair has an arbitrary wave-vector of the center of mass
motion $\textbf{q}$  and an energy $\hbar^2q^2/2m
+\hbar^2\textbf{pq}/m$
  which is the difference between the
conduction electron's energy in the final and initial states,
$(\hbar^2(\textbf{p}+\textbf{q})^2/2m$ and
$\hbar^2\textbf{p}^2/m$, respectively).

The threshold energy  of this excitation at $q=0$ and temperature
$T=0$~K is zero. Therefore, the lower boundary of the continuum
spectrum of the combined excitation consisting of the free exciton
plus an electron -- conduction hole pair  coincides with the energy of the free
exciton. As a consequence, this inelastic scattering process
results in a  homogeneously broadened absorption spectrum with its maximum
shifted toward high energies with increasing free
electron concentration.

The mechanism responsible for the creation of an electron --
conduction hole pair can be either the Coulomb interaction or the
exchange interaction between the electron bound in the exciton and
the free electrons. For the negatively charged trion, which has
two bound electrons coupled in a spin singlet state, the Coulomb
interaction seems to be the only important mechanism of
interaction with free electrons \cite{Klochikhin}. In the case of
the electrically neutral exciton the strong cancellation of the
electron and valence hole self-energy corrections  and vertex
correction decreases the role of the Coulomb interactions in the
scattering processes. This means that the Hartree-Fock exchange
interaction between the electron belonging to the exciton and a
free electron should be considered as the most probable reason for
the broadening of the exciton absorption band.

We consider in detail the behavior of the bound exciton absorption
band as a function of free electron concentration in a quantum
well doped by electrons at low and intermediate concentrations,
and we extrapolate the results up to the concentration at which
the bound exciton state disappears.

We will show in this paper that a second order exchange process
strongly influences the bound exciton state and the optical
spectra of the doped quantum wells, in addition to the nearly
rigid  band gap renormalization  produced by the first order
Hartree-Fock exchange interaction. The second order process
broadens the exciton absorption band and  shifts  the energy of
its maximum.

The paper is organized as follows. In Section II we formulate the
exciton equation in a quasi-two-dimensional quantum well containing free
electrons using the known results for the 3D case. We take into
account in this Section the static screening of the Coulomb
interaction and the effect of phase-space filling. Section III is devoted to
the Hartree-Fock exchange interaction between the electron constituting
the exciton and the free electrons. In Section IV we describe the
calculation of the absorption coefficient as a function of the
free electron concentration and give a comparison with
experimental data.

\section{Exciton in a quantum well}

We consider only temperature $T\rightarrow 0$~K assuming that $k_BT$
is much less than the Fermi energy. Despite the reduced
dimensionality there exists a deep analogy with the 3D Fermi
systems in the methods of calculation of multi-particle effects.

The exciton equation can be formulated by summing the ladder
diagrams for the exciton Green function similarly to the procedure
described by Mahan\cite{Mahan}. Accepting this paper as a guide we
present the exciton equation in a quantum well with finite
concentration of free electrons as
\begin{equation}
-\frac{\nabla^2}{2\mu}\psi_{\lambda}(\overrightarrow{\rho})+\int
d^2\rho'V^{2D}(\rho')
K(\overrightarrow{\rho}-\overrightarrow{\rho'})\psi(\overrightarrow{\rho})=
\end{equation}
$$=E_{\lambda}\psi_{\lambda}(\overrightarrow{\rho}),$$
where $\mu=m_em_h/(m_e+m_h)$ is the reduced mass of the electron
and valence hole, $\overrightarrow{\rho}$ is the two-dimensional
radius-vector between the electron and hole positions in the
exciton. $V^{2D}(\overrightarrow{\rho})$ is the
quasi-two-dimensional Coulomb potential at zero temperature (in
the calculation of which the static screening should be taken into
account), and
\begin{equation}
K(\rho)=-\frac{1}{(2\pi)^2}\int d^2\hat{q}\,\,\,e^{i{\hat{\bf{q}}}
\overrightarrow{\rho}}(1-\Theta({\hat{p}}_F-\hat{q})),
\end{equation}
where ${\hat{p}}_F$ is the Fermi wave vector, $\Theta({\hat{p}}_F-\hat{q})$ is the step-like
theta-function, and
 $(1-\Theta({\hat{p}}_F-\hat{q}))$ is the Fermi distribution
function at zero temperature restricting the integral region in
two-dimensional reciprocal space to $\hat{q}\leq \hat{p}_F$.
 We present $K(\rho)$ as
two items
\begin{equation}K(\rho)=-\delta(\rho)+
\frac{1}{(2\pi)^2}\int d^2\hat{q}\,\,\,e^{i{\hat{\bf{q}}}
\overrightarrow{\rho}}\Theta({\hat{p}}_F-\hat{q}).\end{equation}

After substitution of Eq.(3) into Eq.(1) we obtain two items for the
exciton potential energy. The first item is an attractive
potential $V^{2D}(\overrightarrow{\rho})$ and the second is
a repulsive term that arises  due to phase-space filling
\begin{equation}\delta V^{2D}(\rho)=\frac{1}{(2\pi)^2}\int
d^2\hat{q}\,\int
d^2\overrightarrow{\rho'}\,V^{2D}(\rho')e^{i{\hat{\bf{q}}}(
\overrightarrow{\rho}-\overrightarrow{\rho'})}\Theta({\hat{p}}_F-\hat{q})\end{equation}
The matrix element of this equation responsible for the correction to
the exciton energy is\begin{equation}\int d^2\overrightarrow{\rho}
\psi_{\lambda}(\overrightarrow{\rho})\delta
V^{2D}(\rho)\psi_{\lambda}(\overrightarrow{\rho})= \end{equation}
$$\left[\frac{1}{(2\pi)^2}\right]^2\int
d^2\hat{p}\,d^2\hat{q}\,\psi_{\lambda}(\hat{p})V^{2D}(\hat{q})
\psi_{\lambda}(|\hat{\mathbf{p}}-\hat{\mathbf{q}}|)
\Theta({\hat{p}}_F-\hat{q})$$

In further calculations we find the ground state energy $E_{ex}$
of the Schr\"{o}dinger equation
\begin{equation}
-\frac{\nabla^2}{2\mu}\psi_{\lambda}(\overrightarrow{\rho})+V^{2D}(\rho)
\psi(\overrightarrow{\rho})
=E_{ex}\psi_{ex}(\overrightarrow{\rho}),
\end{equation}
where $V^{2D}(\rho)$ is  the quasi-two dimension Coulomb potential
taking into account the  static screening.
Additionally, we should calculate the phase-space filling
correction according to Eq.(5).

\subsection{ Coulomb potential in a quasi-two-dimensional quantum well}
The  Coulomb and exchange interactions in a quantum well should be
transformed to take into account the fact of the reduced
dimensionality due to the confinement.

The effective Coulomb potential  in the well depends on the
thickness and shape of the quantum well, and on the concentration
of free electrons.

We find the effective Coulomb potential assuming  that  all the
particles occupy their lowest confined state in a quantum well
with infinitely high barriers. In this case the wave functions of all the
particles can be characterized by the quantum well thickness $L$.

As the first step, we obtain an effective "quasi-2D"  Coulomb
potential without static screening
$V_0^{2D}(\overrightarrow{\rho})$ (see Eq.(11) below)by using the
Coulomb interaction in 3D systems as a starting point, similarly
to Ref. [\onlinecite{Klochikhin}]. The Coulomb Hamiltonian in a 3D
system is
\begin{equation}H_{Coul}=\frac{1}{2}\sum_{i,j,\sigma,\sigma'}\int
d^3r_i d^3r_j
V_{Coul}(\mathbf{r}_i-\mathbf{r}_j)|\psi_{\sigma}(\mathbf{r}_i)|^2
|\psi_{\sigma'}(\mathbf{r}_j)|^2,
\end{equation}where $\sigma$ denotes  the spin quantum number, and
\begin{equation}
V_{Coul}(\mathbf{r}_1-\mathbf{r}_2)=\pm\frac{1}{(2\pi)^3}\int d^3q
\frac{4\pi e^2}{\varepsilon
q^2}\exp(i\mathbf{q}(\mathbf{r}_1-\mathbf{r}_2)),
\end{equation} where $\varepsilon$ is the static dielectric constant,
and the $\mathbf{r}_i$, $i=1,2$ are the radius-vectors of the two
interacting particles.  We may express the  $\bf{r}_i$ as
coordinates in an arbitrary plane and perpendicular to this plane
$\mathbf{\bf{r_i}}\rightarrow (\overrightarrow{\rho_i},z)$ and,
similarly, components of the wave-vector $\bf{q}$ in the
reciprocal space  are $\hat{\mathbf{q}}$ and $q_z$. The  signs
$\pm$ are related to the electron-electron and electron -- valence
band hole interactions respectively. In writing Eq. (7) and below, we
are using CGS units.

We exclude the 3D motion in the perpendicular direction by
calculating the matrix elements of exponentials with
wave-functions $\psi(z)=\sqrt{2/L}\cos(\pi z/L)$ assuming that the
wave functions of the exciton and conduction electron --
conduction hole pairs are completely confined within the quantum
well of thickness $L$ restricted by infinite barriers. The Coulomb
vertex contains the product of two matrix elements having the form
\begin{equation}
M_{z,z}=[\exp(izq_z]_{\psi(z),\psi(z)}=\prod_{k=2}^{\infty}[1-(q_zL)^2/(2\pi
k)^2].\end{equation} The integral over $q_z$
\begin{equation}
\int_{0}^{\infty}\frac{dq_z}{2\pi} \frac{4\pi
e^2[M_{z,z}(q_z)]^2}{\varepsilon [\hat{q}^2+q_z^2]}
\end{equation} can be calculated numerically
or represented  with good accuracy by the function $[ \frac{4 e^2
}{\varepsilon \hat{q}}\arctan{\frac{\widetilde{q}_L}{\hat{q}}}]$
where $\widetilde{q}_L=2\pi/\widetilde{L}$ and
$$\frac{L}{\widetilde{L}}=\int_{0}^{\infty}\frac{Ldq_z
}{2\pi} [M_{z,z}(Lq_z/2\pi)]^2\simeq 0.75.$$

As a result we can present the effective quasi-2D Coulomb
interaction without screening as

\begin{equation}
V_0^{2D}(\hat{q})= \frac{4 e^2 }{\varepsilon
\hat{q}}\arctan{\frac{\widetilde{q}_L}{\hat{q}}}.
\end{equation}

The screened effective Coulomb potential in the random phase
approximation has the form

\begin{equation}V^{2D}(\hat{q})= \frac{4 e^2 \arctan{\frac{\widetilde{q}_L}{\hat{q}}}
} {\hat{q}\,\,\varepsilon(\hat{q},0)}\end{equation} where
$\varepsilon(\hat{q},0)$ is the real part of the dielectric
function that takes into account free electrons

\begin{equation}\varepsilon(\hat{q},0)=\varepsilon\left[1-\frac{4 e^2 }{\varepsilon
\hat{q}}\arctan{\frac{\widetilde{q_L}}{\hat{q}}}
Re\chi(\hat{q},0)\right]\end{equation}

In the limit $\omega \rightarrow 0$ the real part of the
dielectric susceptibility  can be presented as
\begin{equation}Re\chi(\hat{q},0)=-\frac{m^*}{\pi
\hbar^2}\{1-\Theta(\hat{q}-2\hat{p}_F)\sqrt{1-(2\hat{p}_F/\hat{q})^2}\}
\end{equation}
And, finally, the screened effective Coulomb potential in the
random phase approximation is
\begin{equation}V^{2D}(\hat{q})= \frac{4 e^2 \arctan{\frac{\widetilde{q_L}}{\hat{q}}}
} {\hat{q}}\times\end{equation} $$
\left\{1+\frac{4}{\pi\hat{q}a_0}\arctan{\frac{\hat{q}_L}{\hat{q}}}\{1-
\Theta(\hat{q}-2\hat{p}_F)\sqrt{1-(2\hat{p}_F/\hat{q})^2}\right\}^{-1},$$
where $a_0=\hbar^2\varepsilon/m^*e^2$ is the radius of the 3D
donor state.

Screening in quasi-2D systems was considered in early publications
by using perturbation theory\cite{Schmitt-Rink-2}. However, new
experimental data\cite{Cox1,Teran} were interpreted as showing
that the exciton binding energy decreases very sharply with an
increase in the concentration of two-dimensional electrons for
CdTe quantum wells and for GaAs/AlGaAs quantum
wells\cite{Kukushkin}.

We have calculated the absolute value of the exciton binding
energy as a function of free electron concentration in the random
phase approximation for 10 nm thick CdTe quantum wells, of the
type studied experimentally in Refs.
\onlinecite{Cox1,Cox2,Cox3,Teran}, assuming infinite-height
barriers. The result is presented as Curve 1 in Fig.1a. The
phase-space filling correction of Eq.(5) is given  in Fig.1a by
Curve 2. The parameters used in our calculations are presented in
Table I.

An effective Rydberg energy of $E_B=18$ meV  for electron
concentration of about $10^{10}$ cm$^{-2}$  was estimated by Teran
et al\cite{Teran} from the energy distance between the ground and
first excited exciton states observed in their PLE spectrum. Such
a value of $E_B$ is obtained in our calculation  at $10^{9}$
cm$^{-2}$ by using the parameters presented in Table 1. We
calculate the quasi-2D Rydberg at zero concentration to be 22 meV.
This is twice the 3D value of 11 meV. The relatively strong
decrease of the effective Rydberg in the low concentration range
from 0 to $10^{9}$ cm$^{-2}$ is explained by the strong screening
of the Coulomb potential at $\hat{q}\le 1/a_0$ in the random phase
approximation as follows from Eq.(15).

From concentrations about $10^{10}$ cm$^{-2}$, the phase-space
filling correction significantly decreases the binding energy.
Extrapolating Curve 2 up to the point where it crosses Curve 1  in
Fig.1a, we can estimate the electron concentration ($\simeq
3\cdot10^{11}$ cm$^{-2}$) above which the bound exciton states
should disappear.

The decrease of the exciton binding energy due to screening and
phase-filling correction is followed by decrease of its oscillator
strength. At zero electron concentration the exciton oscillator
strength is proportional to $[\int
d^2\hat{q}/(2\pi)^2\psi_{ex}(\hat{q})]^2=[\psi_{ex}(\rho=0)]^2\sim
1/a_{ex}^2$ where $a_{ex}$ is the effective exciton radius. For the
oscillator strength at finite electron concentration we have
\begin{equation}A(n_e)= \left[\int \frac{d^2\hat{q}
\Theta(\hat{q}-\hat{p}_F
)\psi_{ex}(\hat{q})}{(2\pi)^2}\right]\psi_{ex}(\rho=0).\end{equation}
In this case two mechanisms decrease the oscillator strength: the
increase of the exciton radius and the phase-space filling
restriction. The major effect is the increase of the exciton
radius resulting from the decrease of the exciton binding energy.
We should mention that some underestimation of the phase space
filling at high electron concentrations occurs in our calculation
of the exciton binding energy because we consider this effect as a
perturbation only.

\begin{table}[b]
\vspace*{6mm} \caption{Parameters used in calculations: electron
and hole effective masses,  the binding energy of the 3D exciton in CdTe
$E_{ex}^{3D}$, the quantum well thickness $t_{QW}$, the
effective exciton radius $a_B^{3D}$, and the static dielectric
constant $\varepsilon$.}
 \vspace {5mm}
\begin{tabular}{|c|c|c|c|c|c|}
  \hline
  $m_e$ & $m_h$& $E_{ex}^{3D}$meV & $t_{QW}$\AA & $a_B^{3D}$\AA & $\varepsilon$ \\
  \hline
  $0.107m_0$ & $0.45m_0$ & 11.0 & 100 & 63 & 10.4  \\
  \hline
\end{tabular}
\end{table}

\begin{figure}[t]
\leavevmode
\centering{\epsfxsize=8cm\epsfbox{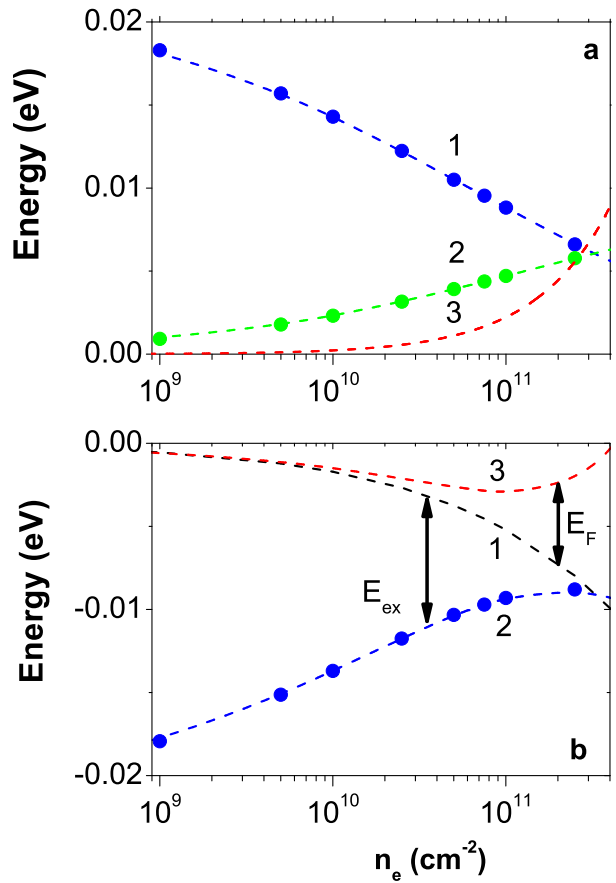}}
 \caption{(Color online) (a) Curve (1) presents the dependence  of the absolute value of the
exciton binding energy on the free electron concentration for a 10
nm wide CdTe quantum well, taking into account only the static
screening. Curve (2) presents the repulsive phase-space filling
correction decreasing the exciton binding energy, and Curve 3 is
the absolute value of the Fermi energy. (b) Curve (1) is the rigid
shift of the band gap $E_G$ corresponding to diagram (a) of Fig.2.
Curve (2) is the exciton binding energy with respect to $E_G$ as a
function of the electron concentration taking into account the
static screening and phase-space filling effects. Curve (3) is the
Fermi energy with respect to $E_G$. }\end{figure}

\section{ Hartree-Fock exchange corrections}

We should say that the inelastic 2D plasmon interaction process
due to Coulomb interaction with free electrons is significantly
weaker for the electro-neutral exciton than for charged particles.
This process does not give a notable additional shift or
broadening of the exciton line, in contrast to its effects for
uncorrelated electrons and holes\cite{Schmitt-Rink-3} or for the
charged trion\cite{Klochikhin}. For the exciton, the cancellation
of the electron and hole self-energy and vertex corrections leads
to a decrease of this interaction. We will show that it is the
Hartree-Fock exchange interaction that can describe the shift and
broadening of the exciton band.

The early calculations of the band gap renormalization in quantum
wells were done in Refs.\cite{Kleinmann,Das Sarma,BauerAndo}

It is convenient to illustrate  the calculations  of the exchange
self-energy corrections by  Feynman diagrams starting from the
corrections to the single electron  interacting with the Fermi sea
of electrons neglecting the electron-valence-band hole exchange
interaction for simplicity.

We consider the first order self-energy correction to the electron
state due to Hartree-Fock exchange interaction, presented in
Fig.2(a), in the framework of the rigid shift approximation,
taking the external wave vector equal to zero. This leads to a
wavevector-independent shift of the conduction band and,
therefore, of the exciton states.

The diagram illustrating the second order correction (Fig.2(b))
can be obtained, for instance, from the Feynman diagrams in Fig.
37(b) of Ref.\cite{Pines} by breaking  one of the electron lines.
 Taking into account that we are interested in broadening of
the exciton created by a $k \approx 0$ photon, we put the value of
the wave-vector of the renormalized particle equal to zero. The
correction to the exciton energy at zero wave-vector of the center
of mass  is given in Fig.2(c). It can be seen that this diagram is
obtained by putting the self-energy  diagram of Fig.2(b) at
$\hat{Q}=0$ into the electron line of the exciton.

Finally, in order to describe the self-energy corrections to the
exciton state, we have to sum the series of multiple scattering
processes. In calculations of the  exciton Green function we
consider only the ground state. This allows us to transform the
series  into a geometrical progression. Therefore, the major part
of the problem is to calculate the self-energy correction to the
ground state.

\begin{figure}
[t] \leavevmode \centering{\epsfxsize=8cm\epsfbox{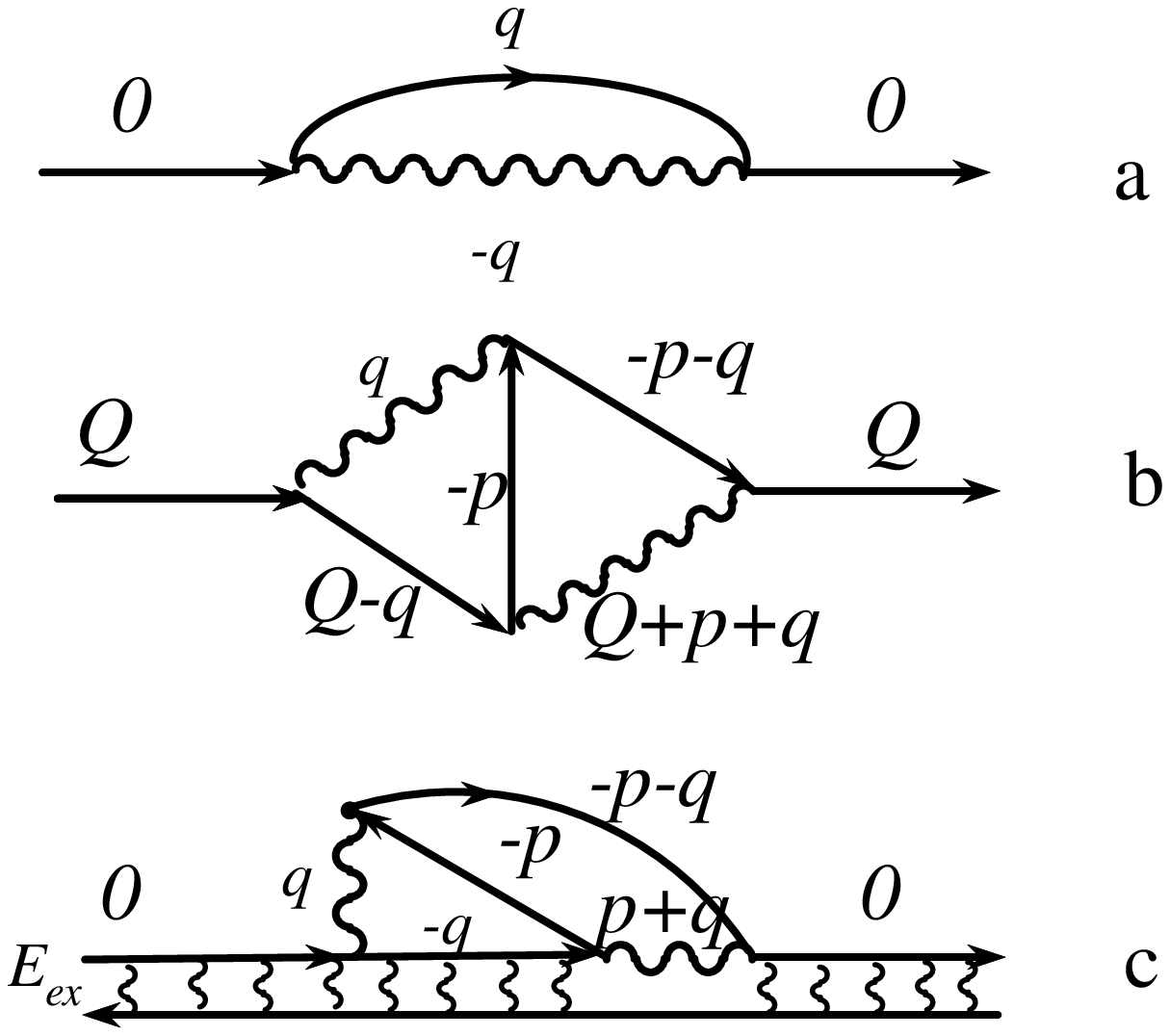}}
 \caption{(a)
Feynman diagram which describes the rigid shift of the electron
energies. (b) The second order exchange self-energy correction to
the electron state. (c) The second order self-energy correction to
the exciton state (double horizontal line connected by the wavy
Coulomb lines) due to exchange interaction, obtained from diagram
(b) at $Q=0$. Solid lines are the electron and hole Green
functions, the wavy lines are the exchange vertexes. }\end{figure}

\subsection{The Hartree-Fock exchange interaction }
First of all we have to express the exchange self-energy
correction to the exciton through the 2D Coulomb-like
$V(q)_{exch}^{2D}$ Hamiltonian taken in the form
$$H_{exch}=\frac{1}{2}\sum_{i,j,\sigma\neq\sigma'}\int d^3r_i d^3r_j
V_{exch}(\mathbf{r}_i-\mathbf{r}_j)\times
$$\begin{equation}\psi_{\sigma}(\mathbf{r}_i)\psi_{\sigma}(\mathbf{r}_j)^*
\psi_{\sigma'}(\mathbf{r}_j)\psi_{\sigma'}(\mathbf{r}_i)^*.
\end{equation}
where the diagram of Fig.2(a) contains the vertex
\begin{equation}
V_{exch}(\mathbf{r}_1-\mathbf{r}_2)=-\frac{1}{(2\pi)^3}\int d^3q
\frac{4\pi e^2}{\varepsilon
q^2}\exp(i\mathbf{q}(\mathbf{r}_1-\mathbf{r}_2)).
\end{equation}
The diagram of Fig.2(c) contains the vertex depending on  the
wave-vector $\mathbf{q}$ and the similar vertex depending on
$(\mathbf{p}+\mathbf{q})$.

The integrals over $z_1$ and $z_2$ for both diagrams we calculate
similarly to Eqs.(9,10). Then the first integral over $dq_z$ gives
the same expression as Eq.(11).
\subsubsection{Rigid shift of the conduction band}
This leads to a rigid shift of the conduction band states $\delta
E_R$
\begin{equation}\delta E_R =\frac{1}{(2\pi)^2}\int d^2\hat{q}
\frac{2 e^2 }{\varepsilon
\hat{q}}\arctan{\frac{\widetilde{q}_L}{\hat{q}}}.\end{equation}
Approximately, the dependence of this equation on the electron
concentration is $\sqrt{n_e}$. The result of the numerical
calculation of the band edge shift is presented in Fig.1b by Curve
(1). The exciton binding energy with respect to the shifted band
edge  is presented in Fig.1(b) by Curve (2).

\begin{figure}
[t] \leavevmode \centering{\epsfxsize=8.5cm\epsfbox{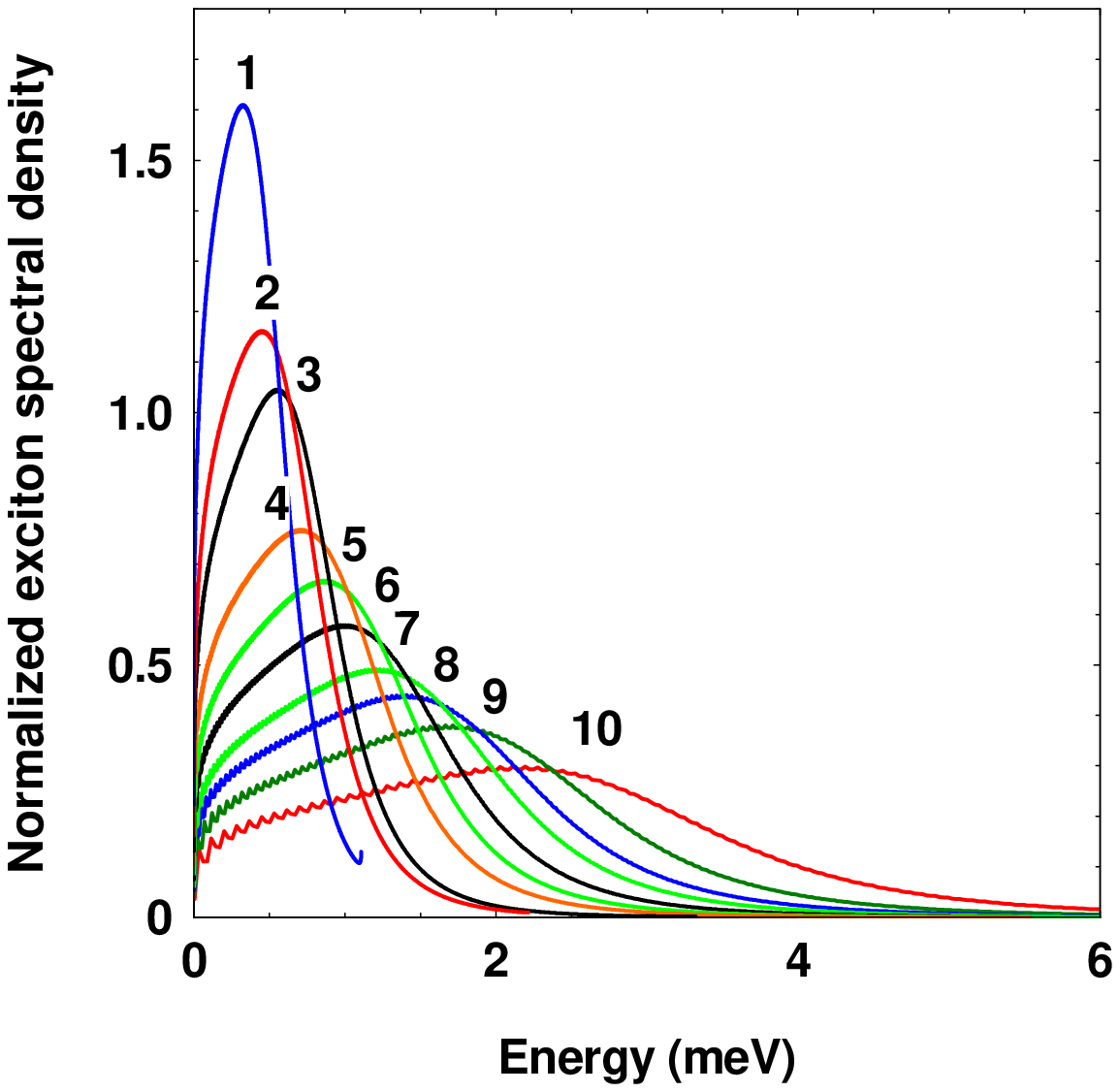}}
 \caption{(Color online)
Broadening and shift of the normalized to unity, spectral density
of the exciton ground state caused by the second order exchange
interaction for ten electron concentrations in a 10 nm CdTe/CdMgTe
quantum well.The zero of energy corresponds to the exciton ground
state at zero temperature and zero electron concentration. The ten
electron concentrations expressed in $10^{10}$  cm$^{-2}$ are
equal to 0.5, 1.0, 1.5, 2.5, 3.75, 5.0, 7.5, 10.0, 15.0, and
25.0.}
\end{figure}

\subsubsection{Shift and broadening of the exciton band
due to the second order correction} The expression corresponding
to the diagram of Fig.2(c) contains two matrix elements over the
internal wave function of the ground state exciton. The first of
them has the form
\begin{equation}\mathcal{M}_{11}(\hat{q})=\int d^2\rho
|\psi_{1}(\rho)|^2\exp(i\hat{\mathbf{q}}\rho)=\end{equation}
$$\left\{\frac{1}{[1+(\hat{q}a_{B}/2)^2]^{3/2}} \right\} $$
and a similar equation can be written   for
$\mathcal{M}_{11}(|\mathbf{\hat{q}}+\mathbf{\hat{p}}|)$.

The structure of the exchange loop in Fig.2(c)
$P_{\hat{q}}^{exch}(\hbar \omega)$ differs from the Coulomb
contribution\cite{Klochikhin} and has the form
$$P_{\mathbf{\hat{q}}}^{exch}(\hbar
\omega)=\int\frac{d^2\hat{q}}{(2\pi)^2}\frac{4 e^2 }{\varepsilon
\hat{q}}\arctan{\frac{\widetilde{q}_L}{\hat{q}}}$$
\begin{equation}\int_{0}^{\infty}\frac{dq_z^1}{2\pi}4\pi
e^2
{M_{zz}^2(q_z^1)}\mathcal{P}_{\mathbf{\hat{q}},q_z^1}^{exch}(\hbar
\omega),
\end{equation}
where the variable $q_z^1=(\mathbf{p}+\mathbf{q})_z$ is
introduced. The function
$\mathcal{P}_{\mathbf{\hat{q}},q_z^1}^{exch}(\hbar \omega)$ is

\begin{equation} \mathcal{P}_{\mathbf{\hat{q}},q_z^1}^{exch}(\hbar
\omega)=\end{equation} $$-\int \frac{d^2p}{2\pi}
\frac{\mathcal{M}_{11}({\hat{q}})\mathcal{M}_{11}(|\mathbf{\hat{q}}+\mathbf{\hat{p}}|)}
{\varepsilon
|\mathbf{\hat{q}}+\mathbf{\hat{p}}|^2+(q_z^1)^2}\eta(\mathbf{p},\mathbf{p}+\mathbf{q},\hbar
\omega).$$ and

\begin{equation}
\eta(\mathbf{\hat{p}},\mathbf{\hat{p}}+\mathbf{\hat{q}},\hbar
\omega)=\end{equation}
$$\left\{\frac{1}{\hbar\omega+i\delta-\omega_{\mathbf{\hat{p}\hat{q}}}}
-\frac{1}{\hbar\omega+
i\delta+\omega_{\mathbf{\hat{p}\hat{q}}}}\right\}.$$

We have neglected the exciton kinetic energy  in the final state
(recoil effect) which is small compared to the energy of the
electron-conduction hole pair.

\begin{figure}
[t] \leavevmode \centering{\epsfxsize=8cm\epsfbox{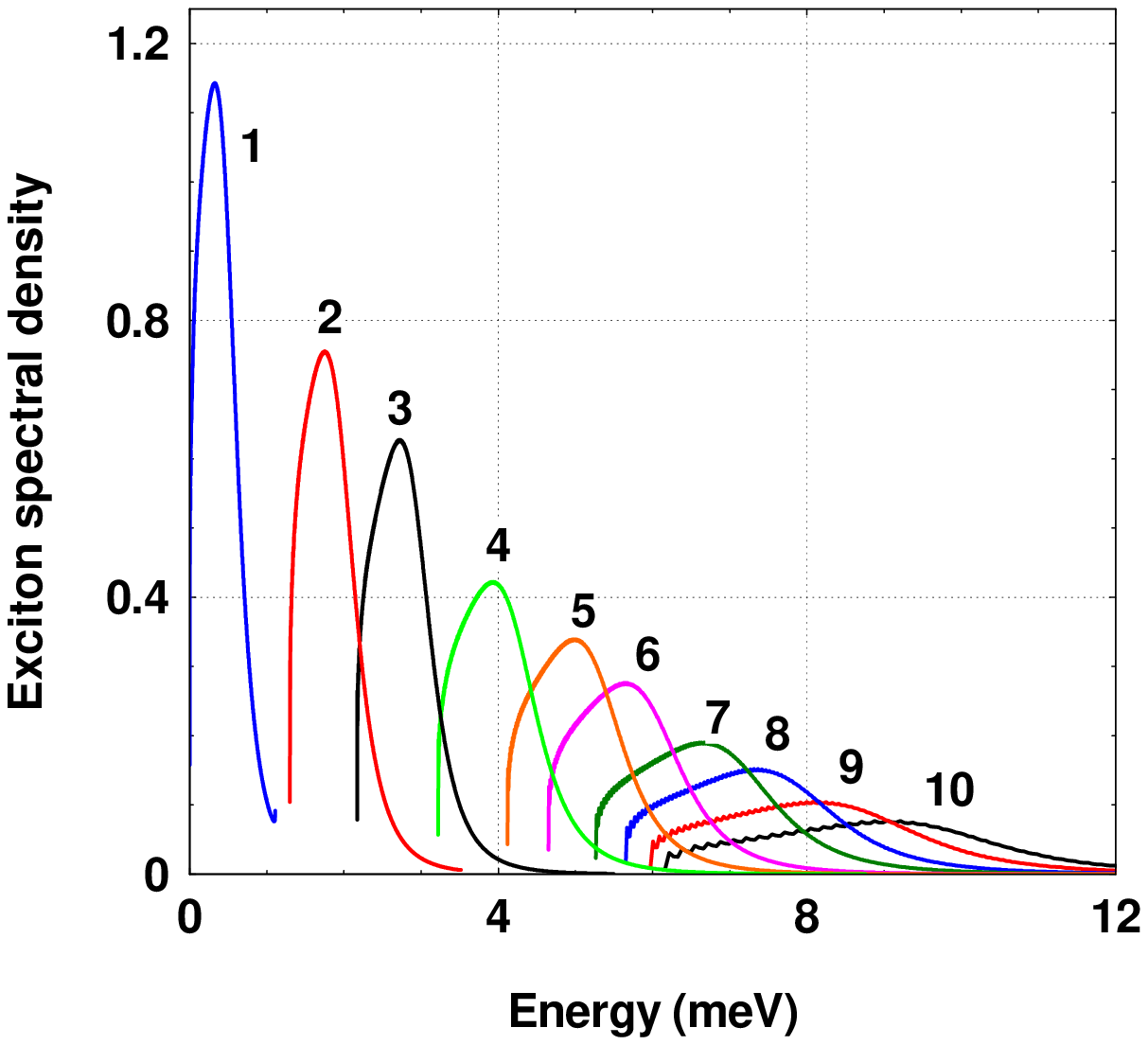}}
 \caption{(Color online)
Spectral density of the exciton ground state for a 10 nm
CdTe/CdMgTe quantum well parameters at increasing electron
concentration. The zero of energy corresponds to the exciton
ground state at zero temperature and  electron concentration equal
to $0.5\cdot10^{10}$ cm$^{-2}$. As compared to Fig. 3, the
threshold energy of each exciton band is shifted upwards~: Here,
in addition to the second order shift and broadening due to
exchange interaction, the static screening, the first-order rigid
shift due to exchange interaction, the phase-space filling effect,
and the decrease of the oscillator strength are all taken into
account. The ten electron concentrations expressed in units of
$10^{10}$ cm$^{-2}$ are equal to 0.5, 1.0, 1.5, 2.5, 3.75, 5.0,
7.5, 10.0, 15.0, and 25.0. }
\end{figure}

The wave-vector dependence of both matrix elements
$\mathcal{M}_{ex}(\hat{q})$ and
$\mathcal{M}_{ex}(|\mathbf{\hat{q}}+\mathbf{\hat{p}}|)$ arises at
values of the order of the reciprocal Bohr radius of the exciton
$1/a_{B}^{2D}$. We will assume this quantity to be essentially
larger than the Fermi wave-vector
 $p_F $ which is the characteristic value for $\mathbf{\hat{p}}$.
Therefore, we approximate the product of these matrix elements by
the expression
$\mathcal{M}_{ex}(\hat{q})\mathcal{M}_{ex}(|\mathbf{\hat{q}}+\mathbf{\hat{p}}|)\approx
\mathcal{M}_{ex}(\hat{q})^2.$
 Taking into account that the integral over $dq_z^1$ is
restricted by $q_z^1\approx q_L=2\pi/L$ we take the function
$\mathcal{P}_{\hat{q},q_z^1}^{exch}(\hbar \omega)$ out of the
integral of Eq.(24)  at $q_z^1= q_L$. After these simplifications
the integral over $dq_z^1$ gives $\widetilde{L}^{-1}$ and the
remaining integral over $d^2p$ in Eq.(22) can be calculated
analytically.

Finally, we obtain the exciton self-energy correction due to
exchange interaction in the form

\begin{equation}\delta E_{ex}(\hbar \omega)=\frac{1}{(2\pi)^2}\int d^2\hat{q}
\left[\frac{4 e^2 \mathcal{M}_{ex}(\hat{q})^2 }{\varepsilon
\hat{q}}\arctan{\frac{\widetilde{q}_L}{\hat{q}}}\right]\times\end{equation}
$$\int d^2\hat{p} \frac{e^2}{\varepsilon \widetilde{L}[|\mathbf{\hat{q}}+\mathbf{\hat{p}}|^2+(q_L)^2]}
\eta(\mathbf{\hat{p}},\mathbf{\hat{p}}+\mathbf{\hat{q}},\hbar
\omega).$$

The broadening of the exciton absorption band and the upward shift
of its peak with respect to the exciton line at zero electron
concentration, resulting from this correction, are shown in Fig.3
for a set of ten electron concentrations ranging up to $n_e =2.5
\cdot 10^{11}$ cm$^{-2}$. The absorption curves have the form:

\begin{equation}\alpha(\omega)\sim\frac{\Gamma(\omega)}
{(\hbar\omega-E_{ex}(n_e=0)-\Delta(\omega))^2+\Gamma^2(\omega)},\end{equation}
where $\Gamma(\omega)$ and $\Delta(\omega)$ are the imaginary and
real parts of the exciton self-energy correction of Eq.(24). For
convenience the absorption band profiles are normalized to unity
in Fig.3. We see that the values of the upward shifts are
comparable with the effect of the rigid downward shift due to the
first order exchange interaction in Fig.1b.

In Fig. 4 we present  the transformed spectra of Fig.3 after the
first-order rigid shift due to exchange interaction, the
phase-space filling effect, as well as  the decrease of the
oscillator strength are all taken into account.
  The shape of the exciton absorption band can be represented in this case as
\begin{equation}\alpha(\omega)=2A(n_e)\frac{\Gamma(\omega)}
{(\hbar\omega-E_{ex}(n_e)-\Delta(\omega))^2+\Gamma^2(\omega)},\end{equation}
where $A(n_e)$ is given by Eq.(16). The position of  the exciton
ground state at zero temperature and  electron concentration equal
to $0.5\cdot10^{10}$ cm$^{-2}$ is taken as the point of reference.

The parameters used in our calculations are presented in Table I.

\begin{figure}
[t] \leavevmode \centering{\epsfxsize=8cm\epsfbox{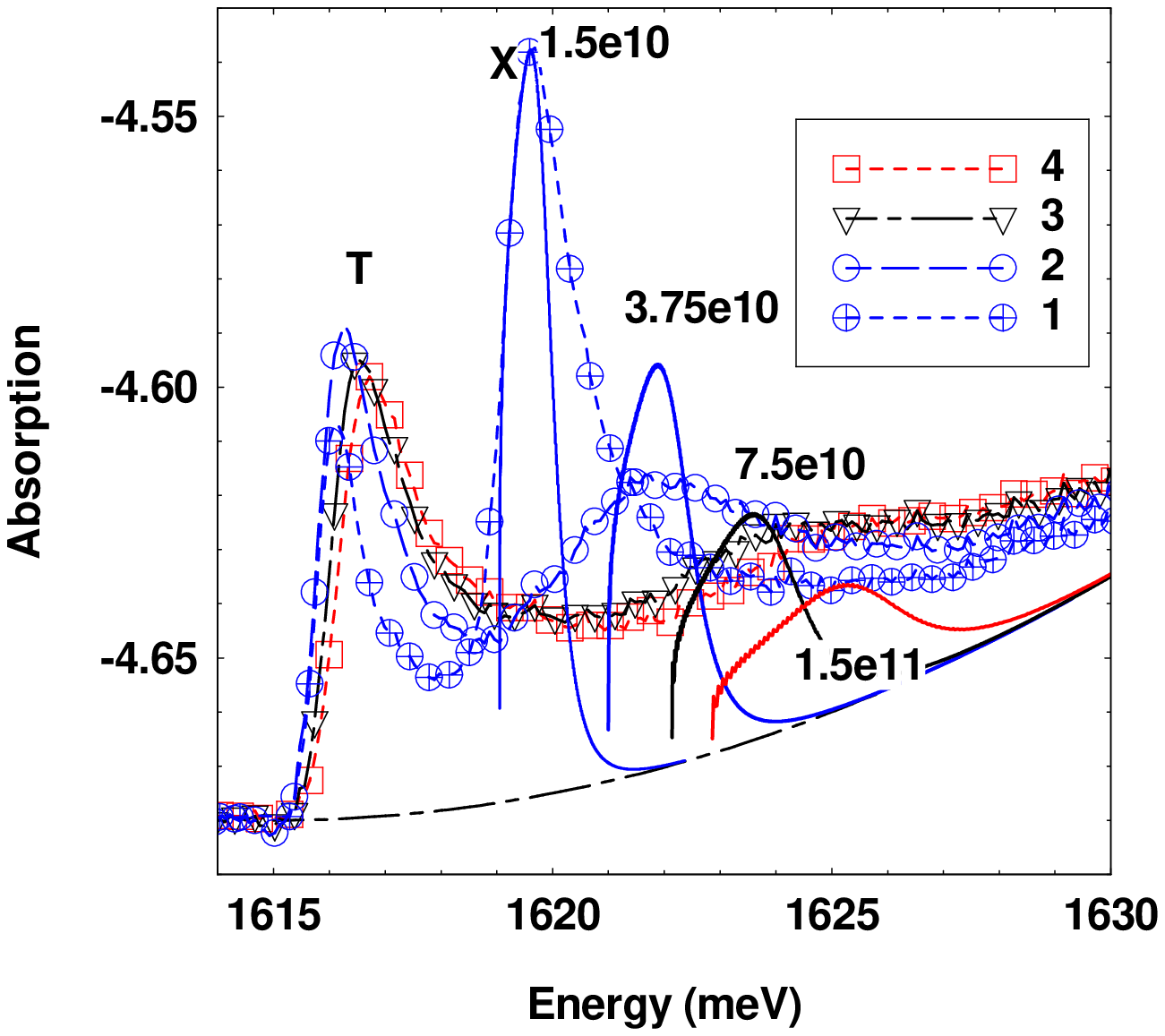}}
\caption{(Color online) Curves (1-4) are the experimental data on
absorption  of a 10 nm CdTe(0.3\%Mn) single quantum well in zero
magnetic field
 at different  free electron
concentrations. A resonance peak (T) corresponds to creation of
the singlet trion appearing below the exciton resonance (X). Solid
curves are the calculated exciton line profiles at free electron
densities of 1.5, 3.75, 7.5, and 15.0 $10^{10}$ cm$^{-2}$. The
dashed curve is a background added to the calculated spectra.
}\end{figure}

\section{Comparison with experimental data}

 We have compared the above theory with optical absorption data at 2K for a
10 nm CdTe/CdMgTe single quantum well at zero magnetic field. This
sample was grown by Molecular Beam epitaxy on a
CdTe$_{0.88}$Zn$_{0.12}$Te substrate. The alloy substrate is
transparent to about 1640 meV, allowing direct measurements of the
optical transmission spectrum in the energy region (1610-1630 meV)
of the heavy-hole exciton in the CdTe quantum well. Also, the
substrate imposes a large biaxial strain on the CdTe well layer,
splitting off light-hole transitions to high energies, which
simplifies the optical spectrum. The upper CdMgTe barrier was
doped with donors so as to give $n_e = 1.5\cdot10^{11}$ cm$^{-2}$
in the well in equilibrium. \cite{FootnoteMn}

Optical data for this sample were provided by Ronald Cox of
N$\acute{e}$el Institute. During the optical transmission
measurements, done using a tungsten lamp source, the electron
concentration could be varied down to a low value by pumping the
sample with blue laser light. The precise dependence of $n_e$ on
pumping light intensity was not known but the minimum
concentration was estimated to be in the range
$(0.2-0.4)\cdot10^{11}$ cm$^{-2}$ as judged from the intensity of
the $X^{-}$ (trion) resonance.

For comparison with this experimental data we will use spectra of
Fig.4 fitting the absolute exciton energy. Results are presented
in Fig.5. As it is seen, the shift of the exciton maximum  agrees
well with experimental data if we take the electron concentration
for the lowest band equal to $1.5\cdot10^{10}$cm$^{-2}$ and for
the highest one to $1.5\cdot10^{11}$cm$^{-2}$. The lowest value
appears to be slightly less than the value estimated in experiment
$(2-4\cdot10^{10}$cm$^{-2}$) while the upper value  agrees well
with experimental data.

However, the experimental broadening values  exceed notably the
calculated values for all the samples. This may be attributed to
the existence of some additional broadening mechanism which was
not taken into account in our calculations.

The existence of an additional mechanism of exciton state decay
might be assumed because the ground exciton state is not the
lowest bound state in the presence of the Fermi sea. At least the
trion state might serve as such a state. Another reason for the
deviation could be the forced simplifications admitted in our
calculations.

\section*{Conclusion}
The shift and broadening of the  exciton absorption bands in
CdTe/CdMgTe  quantum  wells  as a function of the free carrier
concentration at zero magnetic field is investigated. It is shown
that the exciton's shift and broadening behavior is related to the
Hartree-Fock exchange interaction between the electron bound to
the valence band hole and the free carriers. This is in contrast
to the case of the charged trion, where it is the Coulomb
interaction with the free electrons that plays the leading role.
The calculations performed have shown that reasonable agreement
can be obtained with experimental data.

We wish to thank Dr. R. Cox for the experimental data used in the
paper and for many helpful discussions.

\newpage

\removelastskip


\begin{thebibliography}{5}

 \itemsep-2pt

\bibitem{Mahan} G.D.Mahan, Phys. Rev. {\bf 153}, 882 (1967).
\bibitem{Schmitt-Rink-1} S.Schmitt-Rink, C.Ell, and H.Haug  Rhys.Rev. B  {\bf 33}, 1183
(1986).
\bibitem{Schmitt-Rink-2}S.Schmitt-Rink, D.S. Chemla, and D.A.B.Miller,
Advances in Physics {\bf 38}, 89 (1989).
\bibitem{Hawrylak} P.Hawrylak,Phys. Rev. B {\bf 44}, 3821 (1991).
\bibitem{Bauer} G. E. W. Bauer,Phys. Rev. B {\bf 45},  9153 (1992).

\bibitem{Cox1} V. Huard, R. T. Cox, K. Saminadayar, A. Arnoult and S.
Tatarenko,Phys. Rev. Lett. {\bf 84}, 187 (2000).

\bibitem{Cox2} R.T. Cox, V. Huard, C. Bourgognon, K. Saminadayar, S. Tatarenko and R.B. Miller, Acta
Physica Polonica A {\bf 106}, 287 (2004).

\bibitem{Cox3} R.T. Cox, R. B. Miller, K. Saminadayar,T. Baron, Phys.
Rev. B {\bf 69}, 235303 (2004)

\bibitem{Teran} F. J. Teran, Y. Chen, M. Potemski, T. Wojtowicz,
 and G. Karczewski, Phys.Rev. B {\bf 73}, 115336 (2006).

\bibitem{Kukushkin}  S. I. Gubarev,  O. V. Volkov,  V. A. Koval'skii,
D. V. Kulakovskii, I. V. Kukushkin, JETP Lett. {\bf 76}, 575
(2002).

\bibitem{Combescot1} M. Combescot, O. Betbeder-Matibet,  M.A. Dupertuis, Solid State
Communications, {\bf 147},  474 (2008).
\bibitem{Combescot2} M. Combescot, O. Betbeder-Matibet, and F.
Dubin, Eur. Phys. J. B {\bf 42}, 63 (2004).
\bibitem{Combescot3} M. Combescot, O. Betbeder-Matibet,
Solid State Communications {\bf 126}, 687 (2003).
\bibitem{Combescot4} Shiue-Yuan Shiau, Monique Combescot, and Yia-Chung Chang1,
Phys. Rev. B {\bf 86}, 115210 (2012).


\bibitem{Klochikhin} A.A. Klochikhin,V.P. Kochereshko, L. Besombes,
G. Karczewski, T. Wojtowicz, J. Kossut, Phys. Rev. B {\bf 83}, 23
(2011).

\bibitem{Schmitt-Rink-3} S. Schmitt-Rink, C. Ell, S. W. Koch, H. E. Schmidt, and H.
Haug, Solid State Commun. {\bf 52}, 123 (1984).

\bibitem{Kleinmann} D.A.Kleinmann and R.C. Miller, Phys.Rev. B {\bf 32},
2266 (1985).

\bibitem{Das Sarma}S. Das Sarma, R. Jalabert, and S.-R. Eric Yang,
Phys.Rev. B {\bf 41}, 8288 (1990).

\bibitem{BauerAndo} Gerrit E.Bauer and Tsuneya Ando, J. Phys. C:
Solid State Phys. {\bf 19}, 1537 (1986).

\bibitem{Pines}D. Pines and P. Nozieres. The theory of quantum liquids.
Vol.1 Normal Fermi Liquids. W.A.Benjamin,inc. NY,Amsterdam (1966).

\bibitem{FootnoteMn}This CdTe well was doped with 0.3\%Mn, for studying electron
spin-polarization effects in magnetic field. This was of no
concern for the present analysis of zero field spectra: the very
low concentration of Mn does not significantly broaden the exciton
resonance.




\end{thebibliography}
\end{document}